# Iterated stretching and multiple beads-on-a-string phenomena in dilute solutions of highly-extensible flexible polymers

Mónica S. N. Oliveira and Gareth H. McKinley

*Department of Mechanical Engineering, Massachusetts Institute of Technology, Cambridge, MA 02139, US*

The dynamics of elastocapillary thinning in high molecular weight polymer solutions are re-examined using high-speed digital video microscopy. At long times, the evolution of the viscoelastic thread deviates from self-similar exponential decay and competition of elastic, capillary and inertial forces leads to the formation of a periodic array of beads connected by axially-uniform ligaments. This configuration is itself unstable and successive instabilities propagate from the necks connecting the beads and ligaments. This iterated process results in the development of multiple generations of beads in agreement with predictions of Chang et al.[1], although experiments yield a different recursion relation between successive generations. At long times, finite extensibility truncates the iterated instability and axial translation of the bead arrays along the interconnecting threads leads to progressive coalescence before rupture of the fluid column.

It has been known for at least 40 years that the dynamics of capillary thinning and breakup of polymeric jets and threads is substantially different from the equivalent processes in Newtonian fluids[2, 3]. The capillary necking induced by surface tension results in a strong uniaxial stretching flow in the thread, which leads to large molecular elongation and inhibits the finite time singularity associated with breakup in a Newtonian fluid jet[4, 5]. The large viscoelastic stresses ensuing from this stretching can also result in the formation of a characteristic morphology known as a beads-on-a-string structure, in which spherical fluid droplets are interconnected by long thin fluid ligaments. Understanding the distribution of the droplets resulting from the dynamics of this process is important in numerous commercial applications including jet breakup[6], fertilizer spraying[7], roll-coating[8], electrospinning[9] and inkjet printing (for further details see the monograph by Yarin[10]). Similar beads-on-a-string structures have also recently been documented during gravitationally-driven stretching of fluid threads formed from wormlike micellar solutions[11].

The formation of a beads-on-a-string morphology is inherently a nonlinear dynamical process. Classical linear stability analysis shows that a viscoelastic fluid thread is in fact *more* unstable than a Newtonian fluid of equivalent steady-state shear viscosity due to the retardation of the viscoelastic stresses that resist a perturbation of any wavelength[2, 3]. However, the uniaxial extension in the neck near the node of the perturbation results in exponential growth of polymeric stresses in the thinning filament and numerical simulations show the formation of an axially-uniform thread or ligament connecting two droplets [5, 12]. In the spherical beads, the molecules are relaxed and surface tension dominates; whereas in the thin thread, the molecules are highly stretched and viscoelastic stresses dominate.

Simple 'zero-dimensional' analyses of the necking phase of the dynamics (in which axial variations are neglected entirely and the thinning thread is considered to be infinitely long with a spatially-uniform but time-varying radius $R(t)$) show that for quasilinear constitutive equations such as the Maxwell and Oldroyd-B models[13] the exponential growth of the elastic stresses is accompanied by an exponential decay in the filament radius:

$$R(t) = \left(\eta_p R_0 / \lambda \sigma\right)^{1/3} \exp(-t/3\lambda) \qquad (1)$$

where $R_0$ is the initial radius of the filament, $\eta_P$ is the polymer viscosity, $\lambda$ is the relaxation time and $\sigma$ is the surface tension. Direct measurement of this rate of decay using a laser micrometer or a CCD video camera thus enables construction of a capillary-thinning extensional rheometer, which provides quantitative determination of the characteristic fluid relaxation time[14-18].

Recent theoretical analyses and numerical simulations of one-dimensional slender filament models have shown that the profile $R(z,t)$ evolves in a self-similar manner; however, the precise dynamics depend on the relative magnitudes of the inertial, viscous, elastic and capillary terms in the governing equation (see Eggers[19] and Renardy[20] for detailed reviews). Ultimately, this exponential thinning of the viscoelastic fluid thread is truncated by the maximum elongation of the macromolecules. This finite extensibility truncates the exponential stress growth and the thread is then expected to thin linearly in time towards a breakup event with a general form $R(t) \sim (\sigma/\eta_E)(t_c - t)$, where $\eta_E$ is the steady extensional viscosity. The extensional viscosity is much larger than the shear viscosity and depends on the molecular weight of the polymer in solution and the specific form of the nonlinear constitutive model[14, 21, 22]. It is very hard to quantitatively evaluate this property for dilute polymer solutions using other techniques[23]; however, monitoring the slow capillary drainage and ultimate rupture of a necking fluid thread can provide a suitable way of measuring this elusive material function[4, 16, 24].

Recent numerical simulations have also demonstrated the evolution and interaction of multiple beads on a viscoelastic thread. The slow drainage of fluid from the connecting threads into the beads results in small, but non-zero, axial variations in the radii of the slender filaments between the beads and non-zero net tensile forces. Imbalances in the magnitudes of the forces in the threads connected to the two poles of a bead lead to axial drainage and coalescence of beads along a thinning jet[25].

In a very thorough investigation of the linear and nonlinear dynamics of the slender filament equations derived for a finitely extensible nonlinear elastic (FENE) dumbbell model, Chang and coworkers[1] predicted that at long times an additional phenomenon, coined 'iterated stretching', should develop. In this stage of the dynamics, the neck connecting the cylindrical thread to the spherical bead was shown to be unstable to perturbations, which triggered a new instability and an "elastic recoil" close to the neck. This recoil leads to the formation of a new smaller "secondary" spherical drop connected to the primary drop by a thinner cylindrical thread. This new thread thins under the action of capillarity and the necks connecting the thread to the primary drop and new secondary drop may once again become unstable. This hierarchical process can repeat indefinitely, providing the molecules have not reached full extension, leading to multiple generations of beads on strings. Similar iterated instabilities have been predicted numerically and observed experimentally in viscous fluid threads,[26, 27] but do not lead to the formation of beads on a string. In the present Letter, we demonstrate experimentally the analogous phenomenon for the first time in a viscoelastic fluid thread.

To observe this iterated stretching, a number of key physical conditions must be realized[1]. The thinning of a polymer solution described by a FENE constitutive model is controlled by multiple physical parameters that can be combined to give four dimensionless parameters; a Deborah number defined as a ratio of the polymer relaxation time to the Rayleigh time scale for inertio-capillary breakup of a thread of radius $r_0$, $De = \lambda / \sqrt{\rho r_0^3 / \sigma}$; an Ohnesorge number characterizing the importance of viscous effects in the thread, $Oh = \eta_0 / \sqrt{\rho \sigma r_0}$; a solvent viscosity ratio $S = \eta_s / (\eta_s + \eta_p) = \eta_s / \eta_0$ characterizing the relative contributions of the background solvent and the polymer to the total viscosity; and a finite extensibility parameter $L^2$ which characterizes the ratio of the maximum length to equilibrium length of the polymer molecules and which scales with molecular weight of the solute[28]: $L^2 \sim M_w$. Here, $\rho$ is the density of the fluid, $\eta_s$ is the solvent viscosity and $\eta_0$ is the total zero-shear-rate viscosity. All previous numerical predictions and experimental studies of capillary thinning and viscoelastic thread breakup can be represented in different regimes of this four-dimensional parameter space. In particular, Chang et al.[1] demonstrated that for iterated stretching to be observed one requires high Deborah numbers $De \gg 1$, intermediate viscosity ratios ($S \neq 0, 1$), finite fluid inertia $Oh \sim O(1)$ – so that inertial effects lead to rapid growth of the capillary instability and recoil – plus very high finite extensibilities $L^2 \gg 1$ so that the iterated nature of the instability and elastic recoil process is not truncated prematurely by the maximum length of the molecules.

To obtain such values experimentally, in the present experiments we use a high molecular weight water-soluble flexible polymer, poly(ethylene oxide) or PEO, commonly used in drag reduction[29] and viscoelastic jet breakup studies [30, 31]. The specific grade of polymer used (WSR-301) is commercially available and polydisperse, with a molecular weight $\overline{M_w} \approx 3.8 \times 10^6$ g/mol and intrinsic viscosity $[\eta]_0 \approx 1.42 \times 10^3$ cm$^3$/g. The polymer is dissolved at a concentration of 2000 ppm in a mixture of ethylene glycol and water to give a semidilute viscoelastic polymer solution with the viscometric properties shown in Table 1.[32] The high molecular weight and flexibility of the polymer chain indicates an extensibility $L^2 \approx 2.4 \times 10^4$ and our measurements of the fluid relaxation time (described below) indicate that the Deborah number in fluid threads of initial diameter $2r_0 \approx 1.2$mm is $De \approx 127$. Eggers[19] notes that inertial, viscous and capillary effects will all become important in a necking fluid thread (i.e. such that $Oh \sim 1$) on length scales $\ell \sim \eta_0^2/(\rho\sigma)$. For the fluid properties given in Table 1 this corresponds to $\ell \approx 39\mu$m. The iterated necking events will evolve on time scales $t_{Rayleigh} = \sqrt{\rho\ell^3/\sigma} \approx 30\mu$s. We thus use a high-speed digital CMOS video camera (Phantom 5) operating at a frame rate of 1600-1800 fps in conjunction with a high-resolution video microscope lens system (Infinity K2, which confers a spatial resolution of ~2.3μm per pixel) to resolve the late stage dynamics.

In Fig. 1 we show measurements of the global thinning dynamics in a Capillary Breakup Extensional Rheometer (CaBER-1, Cambridge Polymer Group). Initially, the 6mm-diameter plates are separated by a gap $h_i$ = 3mm (Fig. 1(a1)). The liquid bridge confined between the plates is stretched as the top plate moves exponentially ($-50\text{ms} \leq t \leq 0$) to a specified distance $h_0$ = 9.7mm (Fig. 1(a2)). The length of the fluid thread now exceeds the Plateau stability limit and the filament selects its own dynamics so that the viscous, elastic, capillary (and gravitational) forces balance each other. A laser micrometer (Omron ZL4-A), measures the evolution of the midpoint filament diameter, $D(t)$ as the thread thins under the action of capillarity and eventually breaks ($0 \leq t \leq t_f$). A number of different regimes can be discerned in Fig. 1(b). Shortly after the top plate comes to a halt, inertio-capillary oscillations of the hemispherical fluid droplets attached to the end plates occur ($0 \leq t \leq 0.3\text{s}$). These oscillations (with period $T \sim t_{Rayleigh}$) are damped by fluid viscosity and forthwith these regions act as quasi-static fluid reservoirs into which fluid from the necking thread can drain. The oscillations are followed by the rapid development of a central axially-uniform connecting filament of initial diameter $2r_0 \approx 0.4\text{mm}$, which drains very slowly. On these intermediate time and length scales, inertial, viscous and gravitational effects can be neglected and a balance between surface tension and elasticity governs the filament drainage[14]. In this regime, the local extensional rate in the filament is constant and the diameter decays exponentially with time according to Eq. (1). The characteristic relaxation time is extracted by fitting the data to the referred equation, yielding $\lambda$ = 229 ms (solid line in Fig. 1(b)).

At long times $t \geq t_B \approx 1.7\text{s}$ ($t/\lambda \geq 7.5$) the necking thread approaches the characteristic length scale $\ell$ = 39μm discussed above, at which inertial, capillary and viscous effects are all important. The inset video image in Fig. 1(b) shows that a series of regularly-spaced beads form on the viscoelastic fluid. The laser micrometer has a calibrated resolution of 20µm and the signal/noise ratio becomes increasingly poor at these scales. We therefore utilize the high-resolution digital video images for further analysis. A sequence of images showing the formation of the beads near the center of the filament is presented in Fig. 2. The iterated nature of the bead formation process is clear. Following the initial instability and formation of a primary generation of beads, the new interconnecting fluid threads become unstable and form a second and third generation of beads. Using image analysis software, we are able to measure the diameters of the connecting filament at onset of each bead formation event (denoted henceforth $D_N$, for $N = 1, 2,…$) , as well as the bead diameters for each successive generation. The fourth generation is just discernable but hard to quantify as the beads approach the minimum spatial resolution of the image (1 pixel $\approx$ 2.3µm). The data obtained are superimposed onto the laser micrometer measurements in Fig. 3. The formation and growth of each new generation of beads is accompanied by a rapid thinning of the inter-connecting filaments, which deviates from the exponential decay observed in the earlier elastocapillary regime and appears to result in a close to linear decrease in time. These characteristics of the iterated stretching sequence are consistent with the predictions[1]. The analysis of Chang et al.[1] also predicted a recursive relationship between the filament diameters for successive generations, $D_N = f(D_{N-1})$. This is shown in the insert of Fig. 3 for the first four generations formed in six different experimental realizations. The data appear to be fitted to a good approximation by a power law of form:

$$(D_N/D^*) = (D_{N-1}/D^*)^m \qquad (2)$$

A least squares fit of Eq.(2) to the experimental data (solid line) yields fitting parameters $m = 2$ and a characteristic length scale $D^* \approx 44$ µm, which is very close to the Eggers length scale $\ell \approx 39$µm computed a priori. The measured data does not seem to follow the relationship proposed by Chang et al.[1] for generations $N \geq 2$ (shown by the dotted line in the insert):

$$(D_N/D_0) = \sqrt{2}(D_{N-1}/D_0)^{3/2} \qquad (3)$$

where $D_0$ refers to the original filament diameter. The differences are most likely the result of finite extensibility effects as we discuss below.

The source of instability leading to the formation of a beads-on-a-string structure is analyzed in more detail in Fig. 4. For each time, the digitized radial profile $R(z)$ is off-set to show the primary bead (with axial coordinate $z_B(t)$) at the center of the plot. As the primary bead forms, pinching occurs in the necks on each side of the bead ($t-t_B = 50$ms). At this point, the filament at the neck is thinner than in the main thread away from the beads. The fluid gradually recoils ($t-t_B = 100$ms) and feeds a newly developing bead on each side of the primary drop. Meanwhile, the main filament connecting the beads grows thinner and the main bead collapses into an almost spherical shape ($t-t_B = 500$ms). The process of pinching and recoiling can be seen for subsequent generations. For the second generation droplets in Fig. 4, for example, pinching is clear at $t-t_B = 150, 175$ms.

In order to represent the complete spatial and temporal dynamics, from the first stages of droplet formation through coalescence until filament breakage, we process the stream of digital images to construct a 'space-time' diagram[33]. For each frame $i$ (= 1, 2,... 2547) and each axial position $z_j$ ($1 \le j \le 1024$), the average gray-scale intensity along the $x$-axis was calculated after background subtraction. Higher intensities correspond to thicker regions of fluid (*e.g.*, beads), while low intensities correspond to thin fluid elements (thin filament sections). The average intensities were re-scaled from 0 to 1, where 0 represents the thinnest filament present and 1 represents the largest bead in the whole process. The resulting column is added to an array ($i \times j$) to create a space-time diagram (Fig. 5), which captures the evolution of the beads-on-a-string morphology. Initially, the intensity is homogeneous in $z$, showing the existence of a uniform filament. As the first generation of beads forms, we see the appearance of bright bands. Higher generations (and hence smaller) beads appear as progressively lower intensity traces. After about 0.5s ($\approx 2\lambda$), the beads-on-a-string structure is fully established and there is no visible formation of new beads. However, in contrast to the expected rupture event[1], we subsequently observe a new regime in which the fully formed beads migrate axially along the filament. As a result, coalescence between beads of different generations occurs and leads to a coarsening of the pattern. The creation of progressively larger beads is in accord with recent numerical descriptions of draining and merging of beads[25]. The large relief in elastic tension of the fluid thread following each coalescence event is evidenced by the very rapid small-amplitude axial displacements observed (*e.g.* at $t$–$t_B$ = 0.82s and 1.20s). Eventually, the structure has coarsened to a few large beads, the extensibility limit of the polymer is reached and the filament breaks.

Finally, we return to the use of elastocapillary thinning and breakup as an extensional rheometer. A balance of elastic and capillary forces in a axial uniform thread undergoing necking leads to an apparent extensional viscosity that is related to the first derivative of the filament diameter: $\eta_{app} = -\sigma/(dD/dt)$ [16, 34]. The resulting dimensionless Trouton ratio, $Tr = \eta_{app}/\eta_0$ is shown in Fig. 6 as a function of the total Hencky strain, $\varepsilon_H = -2\ln(D(t)/D_p)$. The symbols are obtained by numerical differentiation of the experimental data, and become increasingly noisy as the filament diameter decreases and the discrete resolution of the laser micrometer is approached. To overcome this issue, the diameter profile was also fitted to the expression:

$$D = \left(D_1 + \frac{k_1}{t+t_1}\right)\exp\left(-\frac{t}{3\lambda}\right) - V_2(t-t_2) \qquad (4)$$

with $D_1$, $t_1$, $k_1$, $V_2$, $t_2$ as fitting parameters. This functional form is motivated by the different capillary necking regimes expected theoretically and is able to describe successfully the initial necking and the exponential thinning regimes (via the exponential terms in Eq. (4)), as well as the approach to maximum molecular extensibility and onset of the iterated stretching and drainage regime (the linear term in Eq. (4))[22]. Nonlinear regression yields $D_1$ = 0.067mm, $k_1$ = 0.104mm s, $t_1$ = 0.093s, $V_2$ = 0.1mm s$^{-1}$ and $t_2 = 1.9$s as shown by the solid line in the insert. Analytic differentiation of Eq. (4) results in the solid line shown in the main graph of Fig. 6. The Trouton ratio climbs exponentially and approaches a steady state value at large strains. The asymptotic limit obtained for large strains gives a very large Trouton ratio $Tr_\infty \to \sigma/(\eta_0 V_2) \approx 1.3 \times 10^4$ in good agreement with the theoretical expectation for a dilute solution of FENE dumbbells[13, 23, 25] $Tr_\infty \to 2(1-S)L^2 \approx 4.1 \times 10^4$.

Elastocapillary thinning and breakup thus provides a means of probing the transient extensional response even for very low viscosity – but highly elastic and extensible polymer solutions (*i.e.* viscoelastic fluids with $Oh \ll 1$ but $De \geq 1$ and $L^2 \gg 1$). At late stages of thinning such fluids are prone to iterated instabilities that result in an array of beads-on-a-string and a subsequent slow axial drainage and consolidation phase. Many of the basic features we observe have been described in isolation by existing analyses;[1, 22, 25] however, the interconnected nature of the exponential thinning, iterated instability and coalescence phases has not been analyzed to date. We observe a different recursion relationship (insert in Fig. 3) to that obtained from asymptotic analysis of the Oldroyd-B equation in the limit of infinite $De$. This difference in the observed scaling appears to be a result of additional drainage of the interconnecting elastic threads between successive instability events coupled with the finite extensibility of the PEO chains. The characteristic Hencky strains at which each generation of beads form are shown by the arrows in Fig. 6 and it is clear that the Trouton ratio is no longer climbing exponentially in this regime. It is to be hoped that these final stages of the drainage and breakup of polymer threads will be described by future analytic and numerical studies.

The authors would like to thank Prof. E. S. G. Shaqfeh's research group for providing the polymer solution used in this work. M. S. N. Oliveira wishes to acknowledge Fundação para Ciência e Tecnologia (Portugal) for financial support.

TABLE 1. Properties of the viscoelastic solution used

| $\eta_P$ (mPa s) | $\eta_S$ (mPa s) | $\eta_0$ (mPa s) | S | $\sigma$ (kg s$^{-2}$) | $\rho$ (kg m$^{-3}$) |
|---|---|---|---|---|---|
| 40.32 | 6.77 | 47.09 | 0.144 | 0.0623 | 925 |

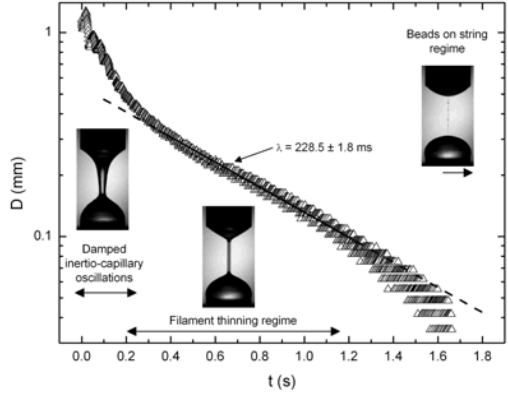

FIG. 1. Capillary Breakup Extensional Rheometry of a PEO solution: (a) CaBER geometry containing a fluid sample (a1) at equilibrium before stretching and (a2) undergoing filament thinning. (b) Evolution of the midpoint filament diameter, $D(t)$ during filament thinning. The symbols represent experimental data obtained using the laser micrometer and the solid line corresponds to the regression using Eq. (1).

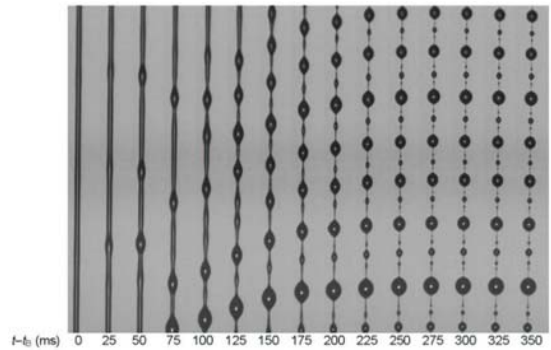

FIG. 2. Sequence of digital video images ($226\times2317\mu m^2$) of the formation and evolution of beads-on-a-string. Development of the first generation of beads-on-a-string occurs at a time denoted $t_B \approx 1.7s$ (first image).

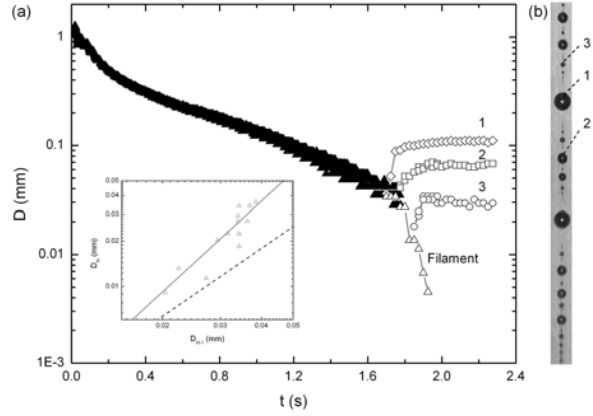

FIG. 3. (a) Evolution of the connecting filament and droplet sizes during late stages of the CaBER experiment. The filled symbols represent filament diameters obtained with the laser micrometer, and the hollow symbols represent measurements obtained using image analysis, corresponding to: Δ filament; ◊ first generation bead; □ second generation bead and ○ third generation bead. The insert shows the recursive relationship between the filament diameters at formation of successive generations of droplets obtained from various experiments. The dashed line corresponds to the Chang et al[1] prediction given by Eq. (3); and the solid line shows the best power law regression with $D^* = 44\mu m$ and $m = 2$ in Eq.(2). (b) Typical image of beads-on-a-string ($138 \times 2317 \mu m^2$) used for image analysis captured at $t-t_B = 500$ ms. The droplets corresponding to those represented in (a) are identified as 1, 2 and 3, for first, second and third generation, respectively.

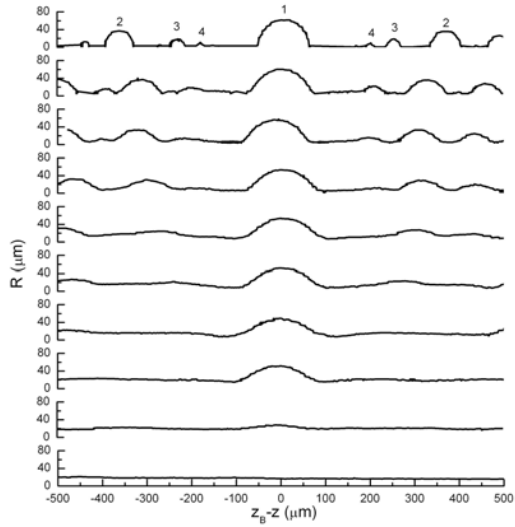

FIG. 4. Progressive evolution of the filament profile for an initially cylindrical fluid filament formed during a CaBER experiment towards a beads-on-a-string morphology. The relative times of each profile are, $t-t_B$: 0, 25, 50, 75, 100, 125, 150, 175, 200, 500 ms, from bottom to top. At each time, the axial position, $z$, has been shifted to show the main bead (with axial coordinate $z_B(t)$) at the center of the plot. The numbers on the top graph classify each droplet in terms of its generation.

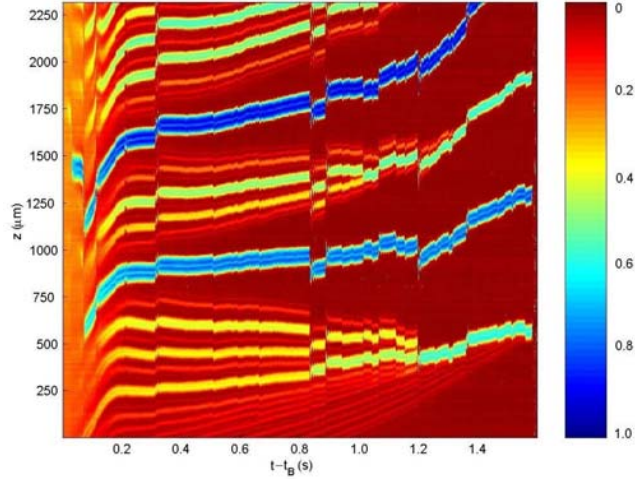

FIG. 5. Space-time diagram of the filament evolution. For each axial position, $z$ and time, $t-t_B$, the colors indicate relative thickness, ranging from zero filament thickness (dark red) to the thickest bead (dark blue).

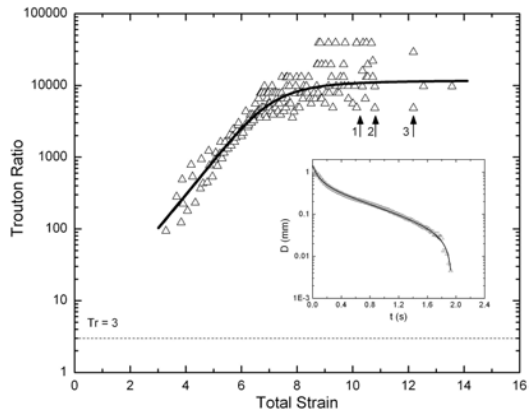

FIG. 6. Apparent dimensionless extensional viscosity obtained from CaBER experiments as a function of the total strain. The symbols are obtained by direct numerical differentiation of the experimental data for the filament diameter, while the solid line is calculated from the analytical derivative of Eq. (4). The labels 1, 2 and 3 indicate the times corresponding to the formation of a first, second and third generation of beads. The insert shows the fit of the measured diameter to Eq. (4).